\documentclass[twocolumn,showpacs,preprintnumbers,amsmath,amssymb, letter]{revtex4}
\usepackage{graphicx}
\usepackage{dcolumn}
\usepackage{bm}

\begin{document}
\title{Small-Worlds, Mazes and Random Walks } 
\author{Bartolo Luque}
\email{bartolo@dma.upm.es} 
\affiliation{ 
Departamento Matem\'atica Aplicada y Estad\'{\i}stica\\ 
Escuela Superior de Ingenieros Aeron\'auticos\\ 
Universidad Polit\'ecnica de Madrid\\ 
Plaza Cardenal Cisneros 3, Madrid 28040, Spain\\}
\author{Octavio Miramontes} 
\email{octavio@fisica.unam.mx}
\affiliation{ 
Departamento de Sistemas Complejos\\ 
Instituto de F\'{\i}sica\\ 
Universidad Nacional Aut\'onoma de M\'exico\\ 
Cd Universitaria, M\'exico 01000 DF, M\'exico} 

\date{\today}
\begin{abstract}
{\normalsize  We  establish  a  relationship between  the  Small-World
behavior  found in  complex  networks  and a  family  of Random  Walks
trajectories using,  as a linking bridge, a  maze iconography.  Simple
methods to generate mazes using  Random Walks are discussed along with
related issues  and it is explained  how to interpret  mazes as graphs
and  loops  as  shortcuts.   Small-World  behavior  was  found  to  be
non-logarithmic but power-law in this model, we discuss the reason for
this peculiar scaling.}
\end{abstract}
\pacs{05.40.Fb, 05.65.+b, 02.50.-r, 02.70.Uu}

\maketitle

Small World  (SW) graphs  efficiently interpolate between  regular and
random  graphs  thanks to  a  small  number  $pN$ of  {\it  shortcuts}
(long-range  connections) which  are superimposed  on a  regular graph
formed by $N$  nodes.  In random graphs, the  mean minimal distance or
diameter   between  all   pairs  of   nodes  in   the   system  scales
logarithmically with  the system  size, while it  does it  linearly in
regular  graphs.  Much  attention  has been  devoted  recently to  the
topological  properties of  such  graphs  and to  the  effects that  a
SW-like connectivity  may have on the properties  of dynamical systems
\cite{watts,bara, Stro, Doro}.

There  is  a  number of  papers  relating  Random  Walks (RW)  and  SW
phenomenon  \cite{Jasch, Lahtinen, Jespersen,  Jespersen2, Jespersen3,
Pandit,Tadic,  manrubia,  miramontes}. In  \cite{Jespersen,Jespersen2}
the authors examine  RW on SW networks, in  particular the probability
of a  random walker  of being at  the original  site at a  later time.
Their interest stem  from the motion of excitons  over polymer chains,
where  steps between  spatially close  sites can  connect  regions far
apart  along  the chemical  backbone.   However  his  model follows  a
standard  SW network  building  \cite{moore}. Later  the same  authors
\cite{Jespersen3},  considering self-avoiding constrains,  assume that
the probability that two sites far apart along the backbone come close
together  in space  is approximately  an  inverse power  law of  their
mutual distance. In  other studies RW have been  employed as dynamical
nodes  to  study  dynamical  SW effects  \cite{manrubia,  miramontes}.
Furthermore and  in another context, RW  on the family  of SW networks
have  been   addressed  where  RW  correspond  to   random  spread  of
information over the network  \cite{Pandit}.  It was demonstrated that
the  average access  time  between nodes  for  a SW  geometry shows  a
crossover  from regular  to random  behavior with  increasing distance
from the starting point of the RW. Average access time is important in
any  Markov  process  and  is  very relevant  for  the  exploring  and
navigation of the  WWW \cite{Tadic} that, as a  scale-free network, it
also shows the SW effect.

\begin{figure}
\includegraphics [scale=0.4]{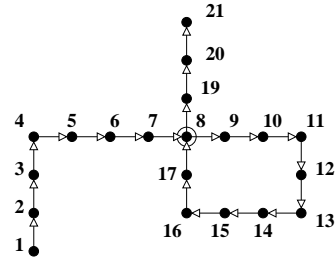}
\caption{\label{LERW} An illustrative example  of short-cut by loop in
the RW path   or    maze:   (a)    The    path   {\bf
1-2-3-4-5-6-7-8-9-10-11-12-13-14-15-16-17-18-19-20-21} of $N=21$ steps
traced by a RW, can be interpreted  as a maze of length $N=21$. (b) To
solve the maze  is: to travel starting at {\bf 1},  and ending at {\bf
21}.  One non-optimal solution is a travel of length $21$. (c) At step
$18$  the  path has  a  self-intersection with  step  8,  a loop  {\bf
8-9-10-11-12-13-14-15-16-17-18}.(d)  We can avoid  this loop  to solve
optimally the  maze. Then, the loop  acts as a short-cut  in the graph
version of the maze: the node  {\bf 8} is connected with the node {\bf
9} and {\bf  19}.(e) To solve efficiently the maze  we use the minimal
distance $L=10$  between nodes {\bf 1} and {\bf
21}: {\bf 1-2-3-4-5-6-7-8-19-20-21}. (f)  Then, the length of the maze
is $N=21$, but using the short-cut, it is solved in $L=11$.}
\end{figure}

\begin{figure}[t]
\includegraphics[scale=0.35]{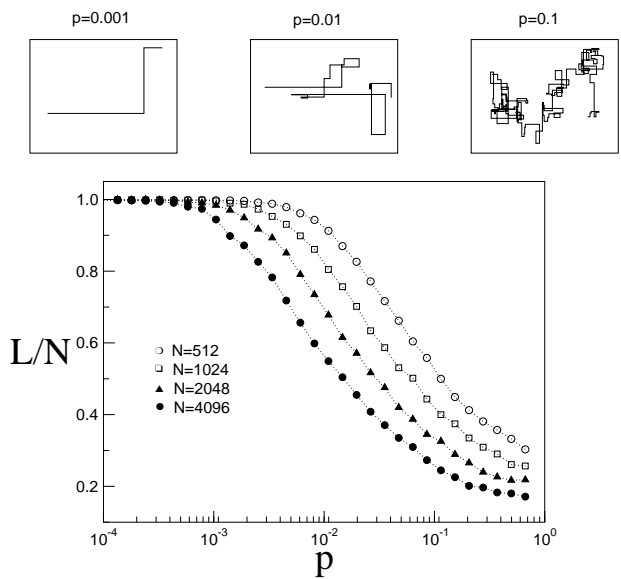}
\caption{\label{ejemplos}  Upper  figures:  Three  examples  of  mazes
generated by RWs with $p=0.001$,  $p=0.01$ , and $p=0.1$ with a proper
choice of  the scale.  Main  figure: The effect  of growth $p$  on the
minimal distance $L$  normalized by $N$ in the  path graph for several
sizes $N$  and log-lineal axes.  For all simulations in  this article,
$31$ points  separated logarithmically between  $p=0.0001$ and $p=2/3$
were  considered.  In  all cases  each  point represents  the mean  of
$1000$ numerical experiments.  The path sizes (or graph  sizes) of the
RW  are:  $N=4096$  (filled  circles), $N=2048$  (triangles),  $N=1024$
(squares), and $N= 512$ (empty  circles).  We notice a size-effect in
$L$  similar to  the  one in  the  classic SW  model  that predicts  a
transition at $p \to 0$ and $N \to \infty$.}
\end{figure}

In this paper we study the properties of two dimensional complex mazes
from the point of view of the  SW theory. For this, it is important to
explore new  methods (such as the  use of the  properties of excitable
media  \cite{Steinbock})  to  find  minimum-length  paths  in  complex
labyrinths.  Navigational methods for solving mazes are widely applied
in  computer sciences for  searching through  data structures  and the
so-called depth-first  search method is  an example \cite{Hofstadter}.
In this work we use  non-reversing RW in two dimensions for generating
mazes  \cite{Pickover}.   We  may  interpret  the  trajectory  of  the
non-reversing  RW in  a plane  as a  directed graph  in  the following
sense: each site reached  at step $i$ by the RW, is  a node labeled by
$i$.  The  nodes are  connected in  the step sequence,  i. e.:  $i \to
i+1$.   But if  a loop  exists,  for example  $i$-step intersect  with
$j$-step  then we connect  $i \to  j+1$.  Therefore  the loops  act as
shortcuts in the graph. In figure  1 we show an example: a path traced
by one of  this non-reversing RW. The path is the  maze and solving it
consist in discovering  the minimum number of steps,  the minimum path
from the first site to the last  site traced by the RW.  As we can see
self-intersections are  no avoided and  so loops are  permitted.  Thus
jumping the loops is a way for quickly reaching the exit.

To generate  a specific  maze we fix  the number  $N$ of steps  of the
non-reversing RW  and the probability  $p\in [0,2/3] $.  In  each step
the RW  will vary  his direction with  probability $p$: at  right with
probability  $p/2$ or  at left  with the  same probability.   And with
probability  $1-p$  the  RW will  not  turn.  In  this manner  we  can
construct  a  variety  of  mazes.  From  $p=0$  that  produces  linear
trajectories to $p = 2/3$ that gives intricate trajectories with equal
probability to continue straight,  turn right or turn left.  Obviously
the  number   of  self-interactions  grows  with  $p$.   Each  time  a
self-interaction occurs, there is a new loop.

In the  upper insets of figure 2  we shown three cases,  with a proper
choice of the scale, for non-reversing RWs with $N=1024$ steps. A very
little value of $p$ such as $p=0.001$ produces a maze without loops. A
value of  $p=0.01$ generates  mazes with a  moderate number  of loops.
Finally, a  value of $p=0.1$ generates very  intricate mazes.  Because
increasing $p$ imply an increase  in the number of loops or equivalent
short-cuts  on a  lineal trajectory  or  regular graph,  we expect  SW
behavior in the model.  We  point out two differences with standard SW
model: here the links are direct and the shortcuts are self-created by
the system dynamics.

In the present work, the SW effect  is used in the sense that the maze
can be solved,  walked from the start to the end,  within few steps on
the average, in spite of the relatively low number of loops present in
the  system.   In  the  main  plot  at figure  2,  the  SW  effect  is
evident. $N$ represents the number of steps taken by the RW and so the
maximum number of steps that takes  going from the start to the end of
the  maze.   If  the  number  of path-intersections  are  regarded  as
shortcuts,  then the  minimal distance  $L$ is  dramatically reduced
with small increments of $p$.

\begin{figure}[t]
\includegraphics[scale=0.3]{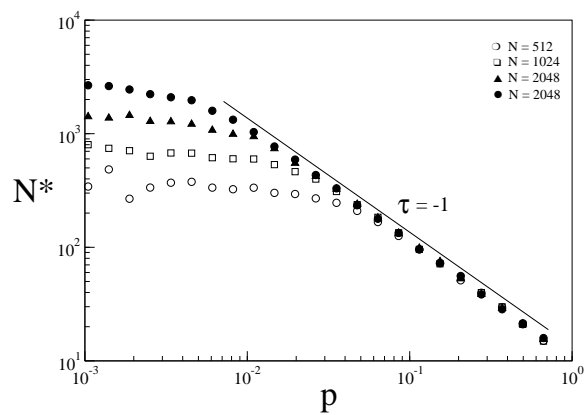}
\caption{\label{tiempos}Log-log  graph  showing  the  scaling  of  the
number of  RW steps $N^*$ that  takes to reach the  first short-cut or
the first loop  as a function of  $p$.  The line with slope  $-1$ is a
guide  for the eye.  The linear  fitting on  the scaling  region gives
$\tau = -1.003 \pm 0.008$.}
\end{figure}

\begin{figure}[t]
\includegraphics[scale=0.3, angle=-90]{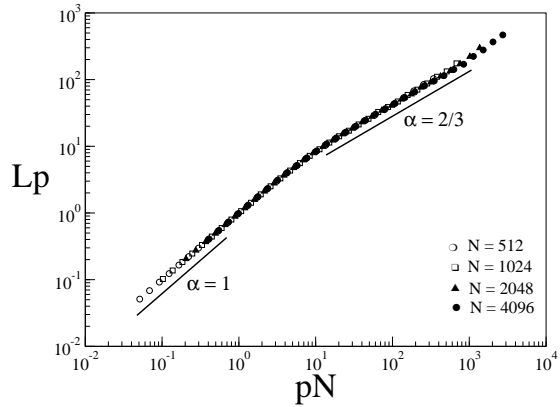}
\caption{\label{colapso1}Log-log  graph showing  the  behavior of  the
scaling function $F_1$. The lines with slope $1$ and slope $2/3$ are a
guide  for the  eye. The  linear fitting  on the
scaling regions give $\alpha =  1.004 \pm 0.007$ and $\alpha = 0.668
\pm 0.003$ for $pN<1$ and $pN>1$ respectively.}
\end{figure}

Similarly  to the SW  standard model,  it is  possible to  determine a
persistence length $N^*$, that is  the mean number of steps needed for
producing the first loop or shortcut.  In Figure 3, it is depicted the
simple dependence regarding $p$: $N^* \sim p^{-1}$, that is equivalent
to the average number of steps  that a RW must take before turning for
the first time.

On Figure  4, we  observe that the  following scaling  relationship is
satisfied:

\begin{equation}\label{1}
  L \sim p^{-1} F_1\bigl(pN\bigr).
\end{equation}

If $p^{-1}$ is proportional to the mean number of RW steps for producing
one loop, $F_1\bigl(pN\bigr)$ can be interpreted as the mean number of
loops  in  a  system  with   size  $N$  and  probability  $p$  in  its
building. The scaling function $F_1(x)$ behaves linearly for values of
$x<1$. That  is, while $pN$ -the  average number of  turnings- is less
than  one there are  no shortcuts  in the  maze and  the value  of $N$
matches $L$.  When  $x>1$ the presence of shortcuts  has an impact and
the distance  $L$ is  reduced following a  power-law with  an exponent
value $\alpha$ equal to $2/3$ when the size of the maze is increase.

We depict in Figure 5 a collapse plot of the scaling function $F_2(x)$:
 
\begin{equation}\label{2}
  L \sim N F_2\bigl(pN\bigr).
\end{equation}

\begin{figure}[t]
\includegraphics[scale=0.3, angle=-90]{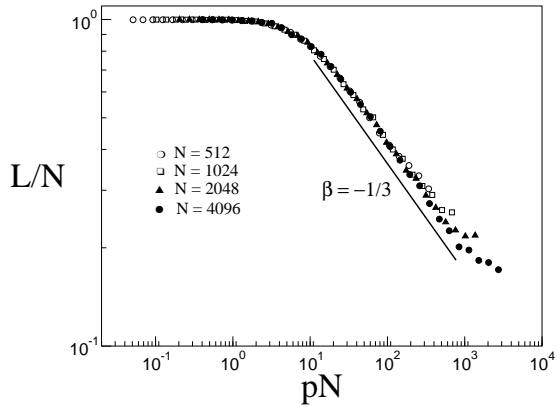}
\caption{\label{cola2}Collapse  of  the  scaling function  $F_2$.  The
line with slope $-1/3$ is a guide for the eye. The linear fitting 
on the  scaling region $Np>1$ gives $\beta = -0.332 \pm 0.003$.}
\end{figure}

$F_2(x)$ may be  interpreted as the reduction factor  in the start-end
maze distance  as a function of the  size and the turning  rate on the
system given by $p$. Again, for  values of $x<1$, $L$ and $N$ take the 
same value.  However, as $N$ or $p$ are increase in such a way
that  $pN >  1$  the ratio  $L/N$ decreases  as  a power  law with  an
exponent value $\beta$ equal to $-1/3$.

Both results confirm that:
\begin{equation}\label{3}
L(N,p) \sim N,
\end{equation}
for no loops or $pN<1$. On the other hand:
\begin{equation}\label{4}
L(N,p) \sim p^{-1/3}N^{2/3},
\end{equation}
for $pN>1$,  when loops  appears.  This result  is particular  for the
present  model, because  in the  classic  SW model  the mean  distance
scales as $log  N$ and in the present system it  scales as a power-law
with exponent $2/3$.

Why the  model presents this SW  effect in a power-law  form?  For our
numerical experiments  we have computed the Euclidian  distance $R$ in
the two dimensional  plane of the line linking the  start with the end
of  the maze,  the  classical end-to-end  distance  for a  RW.  For  a
typical RW in two dimensions $R$  scales with $N$ as $R \sim N^{1/2}$.
In figure 6 it is shown how  $pR$ scales with $pN$ as a power-law with
exponent $1/2$:
\begin{equation}\label{5}
  pR \sim (pN)^{1/2}.
\end{equation}\\

\begin{figure}[h]
\includegraphics[scale=0.3]{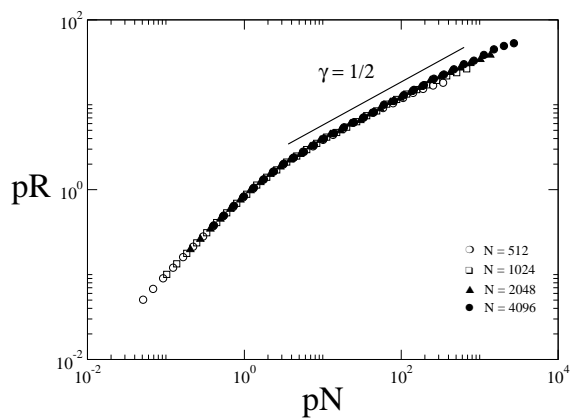}
\caption{\label{R}Scaling relation between the end-to-end distance $R$
and $N$,  the number of  the RW steps  rescaled both by $p$.  The line
with  slope  $1/2$ is  a  guide for  the  eye.   The scaling  exponent
obtained  by fitting  on  the  scaling region  is  $\gamma =0.499  \pm
0.004$.}
\end{figure}

This result  shows that the maze model  acts as a typical  RW model of
$N$ steps in two dimensions where $N$ and $R$ are scaled by $p$.

Self-Avoiding   Random   Walks  (SAW)   are   RW  where   trajectories
self-intersections are  avoided.  In our model once  the RW trajectory
is  finished, deleting the  loops produces  a SAW  of length  $L$. Its
known that  in a typical SAW model  in two dimensions of  $L$ steps we
can  expected $R  \sim L^{3/4}$  \cite{DeGennes} as  scaling relation.
This imply in  our model, that after rescaled $R$ and  $L$ as $pR$ and
$pL$:

\begin{equation}\label{6}
pR \sim (pL)^{3/4}.
\end{equation}

In figure (7) the above scaling relationship is confirmed.

Then, after  combining equations (5)  and (6) we can  recover equation
(4).   The SW  effect  in the  two  dimensional model  as a  power-law
scaling between $N$ and $L$ with  exponent $2/3$ for values $pN$, is a
direct consequence of the well-known scaling relations in classical RW
and SAW models.

\begin{figure}
\includegraphics[scale=0.3]{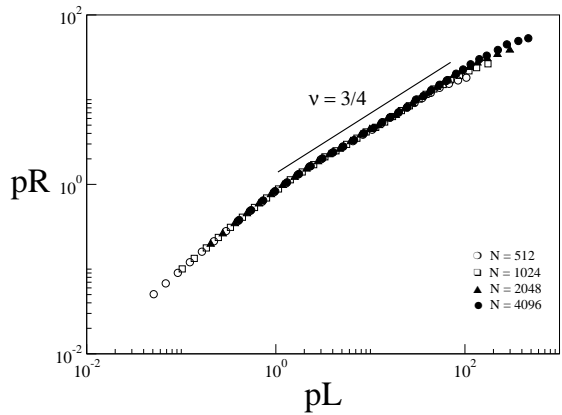}
\caption{\label{R-L} Scaling relation  between the end-to-end distance
$R$ and $L$ scaled  both by $p$. The line with slope  $3/4$ is a guide
for  the eye.  Scaling  exponent  obtained on  the  scaling region  by
fitting is $\nu = 0.75 \pm 0.01$.}
\end{figure}

\section*{ACKNOWLEDGMENTS}

We would like to thank Ugo Bastolla and Denis Boyer for their valuable
opinions. OM has been supported  by CONACYT (32453-E and G32723-E) and
DGAPA-UNAM (IN-111000) and BL by CICYT BFM2002-01812.

\end{document}